# HETEROGENEITY IN SHORT GAMMA-RAY BURSTS

Jay P. Norris[1], Neil Gehrels[2], and Jeffrey D. Scargle[3]

[1] Physics and Astronomy Department,
University of Denver, Denver CO 80208

[2] Astroparticle Physics Laboratory,
NASA/Goddard Space Flight Center, Greenbelt, MD 20771

[3] Space Science and Astrobiology Division,
NASA/Ames Research Center, Moffett Field, CA 94035-1000.

## ABSTRACT

We analyze the *Swift*/BAT sample of short gamma-ray bursts, using an objective Bayesian Block procedure to extract temporal descriptors of the bursts' initial pulse complexes (IPCs). The sample comprises 12 and 41 bursts with and without extended emission (EE) components, respectively. IPCs of non-EE bursts are dominated by single pulse structures, while EE bursts tend to have two or more pulse structures. The medians of characteristic timescales – durations, pulse structure widths, and peak intervals – for EE bursts are factors of ~ 2–3 longer than for non-EE bursts. A trend previously reported by Hakkila and colleagues unifying long and short bursts – the anti-correlation of pulse intensity and width – continues in the two short burst groups, with non-EE bursts extending to more intense, narrower pulses. In addition we find that preceding and succeeding pulse intensities are anti-correlated with pulse interval. We also examine the short burst X-ray afterglows as observed by the *Swift*/XRT. The median flux of the initial XRT detections for EE bursts (~ $6 \times 10^{-10}$ erg cm$^{-2}$ s$^{-1}$) is $\gtrsim 20 \times$ brighter than for non-EE bursts, and the median X-ray afterglow duration for EE bursts (~ 60,000 s) is ~ $30 \times$ longer than for non-EE bursts.

The tendency for EE bursts toward longer prompt-emission timescales and higher initial X-ray afterglow fluxes implies larger energy injections powering the afterglows. The longer-lasting X-ray afterglows of EE bursts may suggest that a significant fraction explode into more dense environments than non-EE bursts, or that the sometimes-dominant EE component efficiently powers the afterglow. Combined, these results favor different progenitors for EE and non-EE short bursts.

Subject headings: gamma-ray burst: general

# 1. INTRODUCTION

Short gamma-ray bursts (GRBs) manifest an initial pulse complex (IPC) lasting $\sim 10^{-2} - 1$ s in the low-energy gamma-ray regime, $\sim 10$ keV – 1 MeV (Mazets et al. 1981; Norris et al. 1984; Kouveliotou et al. 1993; Meegan et al. 1996; Nakar 2007). Time profiles of IPCs often contain one or just a few narrow pulses. These pulses or pulse structures (clusters of pulses) in short GRBs have little information content due to their brevity ($\sim 10$–$20 \times$ shorter than pulses in long GRBs) and therefore their tendency to relatively low fluence (Norris, Scargle & Bonnell 2001). Approximately one-tenth of the Swift sample are short bursts, their optical afterglows are detected less frequently than those of long bursts, and when detected the afterglow position is not always unambiguously associated with one galaxy. Therefore compared to long bursts, less is known definitively about short bursts' host galaxies, progenitor environments, and redshifts (Kann et al. 2008; Berger 2010; Fong, Berger & Fox 2010), although it is clear that their hosts and positions within hosts favor regions of lower star formation and thus a population of older progenitors such as NS-NS mergers (Berger 2009).

Recently it has been discovered that the IPCs of up to $\sim$ one fourth of short GRBs are followed by a softer, extended emission (EE) component lasting $\sim 100$ s, as evidenced in the *Compton*/BATSE, Konus, HETE-II, *Swift*/BAT, and SPI-ACS/INTEGRAL samples (Mazets et al. 2002; Fredericks et al. 2004; Barthelmy et al. 2005; Villasenor et al. 2005; Gehrels et al. 2006; Norris & Bonnell 2006: NB06; Norris & Gehrels 2008; Minaev, Pozanenko & Loznikov 2010). The IPCs of bursts with and without EE tend to exhibit negligible or very short spectral lags ($\sim$ milliseconds) between low-energy gamma-ray bands, which could suggest a unified population. However, some bright BATSE and BAT short bursts have intensity ratios, EE upper limit : peak IPC, of $\lesssim 10^{-4}$, arguing for complete absence of the EE component, and therefore the possibility that different progenitors might give rise to EE and non-EE short bursts (NB06). Norris, Gehrels & Scargle (2010: NGS10) calibrated the BAT sensitivity to an EE component, and reported that EE should have been detected in $\sim$ half of the BAT bursts if it were present, compared to the fraction with detected EE of $\sim$ one fourth. Thus half to three fourths of BAT short bursts appear to be truly short – without an EE component. The NGS10 result does not address the question of whether there is one short GRB progenitor type for which an EE turn-on threshold mechanism operates, or possibly two (or more) short GRB progenitor types.

Figure 1 illustrates the duration distribution for 327 BATSE short bursts where their time-tagged-event (TTE) data contain the whole IPC. These durations – comparable to the usual $T_{90}$ – were measured using the same Bayesian Block procedure described in the next section for *Swift*/BAT bursts. The durations span $10^{-2} - 2$ s with apparent modes near $\sim 60$ ms and $\sim 250$ ms (the robustness of the double-mode appearance varies depending on choice of the histogram binning scale). Also plotted are durations for the eight BATSE short bursts (diamonds) with EE that were detailed in Norris & Bonnell (2006). It appears striking that these eight bursts occupy the long part of the distribution, all to the right of the 250 ms mode. While brightness selection effects may be operating – e.g., perhaps EE bursts with long IPCs tended to be more readily identified than those with short IPCs – we did not note such an effect in the procedures of Norris & Bonnell (2006). Thus we are led to conjecture that short bursts with EE may tend to have temporal structure with longer timescales than those without EE.

Previous analyses have used less than optimal data to characterize short burst IPCs. The duration distributions in Figure 5 of NGS10 were measured using *Swift*/BAT's mask-tagged data type with 64 ms resolution. While there is a hint in the figure that non-EE bursts tend to be shorter than EE bursts, 64 ms bins are too long for accurate measurements on the short end of the distribution, and the mask-tagged data contains only ~ 30% of the signal present in the BAT event data type (Fenimore et al. 2003). The EE and non-EE BATSE bursts represented in Figure 6 of NGS10 (adapted from Figure 2, Norris & Gehrels 2008) were segregated only in a statistical sense as being probable candidates for either group; variability of the all-sky background that obtained for BATSE made individual definitive classification for most bursts problematic. Comparative analyses of the optical and X-ray afterglows of long and short bursts have been reported previously (Kann et al. 2008; Nysewander, Fruchter & Pe'er 2009; Berger 2010), but not with an eye for exploring possible differences between EE and non-EE short bursts.

In this work we use the BAT event data in an objective Bayesian Block (BB) methodology (Scargle 1998) to analyze the time profiles of short burst IPCs. In Section 2 we describe the BAT sample and details of the analysis procedures. Our results of the IPC analysis – measurements of temporal descriptors for the two groups – are presented in Section 3. We then examine the *Swift*/X-Ray Telescope (XRT) light curves of short bursts in Section 4. Ramifications for short GRB progenitors and connections to other avenues for classification are discussed in Section 5.

## 2. ANALYSIS OF INITIAL PULSE COMPLEXES

The short burst sample analyzed here includes the sample of NGS10 – 12 with EE and 39 without EE – augmented by three recent bursts, GRBs 100625A, 100702A, and 100816A (none exhibited EE). One burst in the NGS10 sample (GRB 050906) was eliminated since the IPC was not detected by the BB algorithm, leaving a total of 53 bursts.

We used the BAT raw event data for the IPC analysis. The arrival time of each photon is recorded with native resolution of 100 μs (two photons may have the same time assigned), and the deposited energy is recorded with 0.1 keV precision. Each time series was prepared by summing over events with energies in the range 15–350 keV, and then selecting those events in the interval within [−10, +12] s of the time of burst peak intensity. The time series were then binned to 10 ms and 5 ms resolution and the whole analysis was performed with both resolutions. At 5-ms binning the additional information afforded over 10-ms binning nears the point of diminishing returns, principally because most bursts are not sufficiently bright to merit the analysis described here on shorter timescales. The results we report are for 5-ms binning. The constructed 22-s time series thus contain 4400 bins, with ~ 10 s of background before and after the region containing the IPC. The pre and post IPC background levels for each time series, as represented by the BB algorithm differed – with one exception – by < 2.5%. The ratio of pre to post background levels was distributed approximately equally about unity for both the EE and non-EE groups. For the exception, GRB 060614, we shortened the time series to 15 s to eliminate onset of the EE component.

We utilize a BB algorithm to realize representations of the IPC time series from which we derive descriptors of temporal features. The algorithm is objective and automatic, does not require pre-defined bins or thresholds, and optimally reveals signal but suppresses noise, based on a sound statistical procedure. The BB compression of a continuous signal into discontinuous blocks, despite appearances, preserves most or all of the information present in the data. Jackson et al. (2003) gives the proof of the theorem for optimal BB partitioning.

While the BB algorithm accommodates "unbinned" event data (in practice single events are always recorded on a discretized bin scale, however short), in its present formulation the computation time goes as $N_{events}^2$ (or $N_{bins}^2$). For the 22-s IPC time series the median background level ranged ~ 5000–9000 counts s$^{-1}$, which translates to ~ 1–2 × 10$^5$ events. For this reason as described above, we elected to bin the event data into 5-ms bins to realize reasonable computation times. Scargle et al. (2011) develops cost (or fitness) functions for several data types and describes how to calibrate the one (apparently) free parameter, ncp$_{prior}$, the adopted prior on the number of blocks which affects the relative probability of any partition into blocks, e.g., the decision to divide one block into two. For binned Poisson-distributed data the maximum likelihood cost function is

$$\log L_{max(binned)} = N^{(k)} [\log N^{(k)} - \log M^{(k)}] - ncp_{prior} \qquad (1)$$

where the $M^{(k)}$ and $N^{(k)}$ are the interval sizes and counts, respectively, for the k$^{th}$ trial block. We set the value for ncp$_{prior}$ by performing simulations of 4000-bin time series for a one-block (constant mean) model with Poisson-distributed data. For ncp$_{prior}$ = 7.5, the value adopted here, one false positive (extra) block occurs once in ~ 100 trials (cf.: NGS10 appendix). With this low rate of false positive blocks we would be assured that the BB representations of the IPCs do not contain extraneous temporal structure. However, we noticed the occurrence of (exclusively) positive spikes in the BB representations, almost always one bin wide and in random positions – not associated with burst structure – essentially extra-Poisson fluctuations ranging ~ 20–50 counts per 5-ms bin above background level. When binned to 5 ms, the spike frequency is often ~ twice per 4400-bin time series. Upon examination of several cases using the event data at the native 100 μs resolution, we found that the spikes consist of a few tens of events with exactly the same time tag, suggestive of an instrumental origin. (Of course, some events with equivalent time tags are real, with higher probabilities for occurrence in burst intervals; hence it seems injudicious to eliminate arbitrarily all such events.) We addressed these extraneous spikes, replacing their values in the 5-ms bins with the average of counts in four nearby bins, and then rerunning the BB algorithm. The result of this procedure is that the spike block, and preceding and succeeding blocks become unified, with the time interval in question then represented by one block (invariably near background level).

The BB representation then contains the information necessary to make estimates of significant temporal features in the IPC time series. We define the background level ($I_{bck}$) as the average of the intensities of the first and last blocks in the representation (except for GRB 060614, for which we used only the first block). The mid time of a block which is a local maximum – with at least one less intense block adjacent on both sides – is chosen as the peak time ($t_{pk}$) of a pulse structure. The number of local maxima is $N_{pk}$. The time between two peak times defines a peak interval ($\tau_{intv}$). The mid time of the least intense block between two local maxima peaks is chosen as a valley time ($t_{val}$). The intensity of a pulse structure ($I_{pul}$) is defined as its intensity at $t_{pk}$ minus $I_{bck}$; similarly for valleys, $I_{val}$ = intensity at $t_{val}$ minus $I_{bck}$.

The prescription for estimation of pulse structure widths is more complicated, as governed by the relative depths of valleys on either side of the structure. We define the width of a pulse structure, $\tau_{wid}$, as the time between rise and decay points where the rising or decaying structure intensity is greater than $I_{1/e} = (1/e) \times I_{pul}$. Case 1: If any block in the valley region before (after) the peak is below $I_{1/e}$, the end (start) time of that block is defined the start of the $\tau_{wid}$ interval. Case 2: If all blocks in the valley are above $I_{1/e}$ then a portion ($\Delta \tau_{wid}$) of the lowest block interval ($\tau_{low}$) in the valley region being considered is accumulated to $\tau_{wid}$ – along with all blocks of higher intensity in the pulse structure – according to the following prescriptions. For the rising part of the structure,

$$\Delta \tau_{widris} = f_{ris} \, \tau_{low} \, R_{pk} \qquad (2)$$

with $R_{pk} = I_{pul} / (I_{pul} + I_{pulpre})$

where $I_{pul}$ is the intensity of the pulse structure and $I_{pulpre}$ that of the preceding structure. Analogously for the decaying part,

$$\Delta\tau_{widdec} = f_{dec}\,\tau_{low}\,R_{pk} \qquad\qquad (3)$$

$$\text{with } R_{pk} = I_{pul} \,/\, (I_{pul} + I_{pulpost})\,.$$

The fraction values $f_{ris} = 1/3$ and $f_{dec} = 2/3$ are assigned to reflect the tendency of pulses to have longer decays than rises. The sense of the cofactor $R_{pk}$ is that for two structures which share a valley, the more intense structure receives a part of $\tau_{low}$ in proportion to the structure's fraction of the summed intensities. The sparsity of blocks (usually ~ 1 or 2) between peak blocks argues that further refinement in the definition of pulse structure width (e.g., fitting models) would tend toward over interpretation of the BB representation. The duration ($\tau_{dur}$) is defined – similarly to Case 1 for widths – as the interval from the $I_{1/e}$ times of the first (rise) and last (decay) peaks in the time series with intensities $> I_{bck}$. Also, the $\tau_{dur}$ start and stop times serve as surrogates for the first and last valley times. $\tau_{dur}$ is comparable to the usual $T_{90}$ measure. In the next section, we compare the distributions of these descriptors for the EE and non-EE bursts.

## 3. TEMPORAL DESCRIPTORS OF INITIAL PULSE COMPLEXES

We list in Table 1 the GRB date and derived average properties of short burst IPCs: number of peaks ($N_{pk}$), intensity above background of the brightest pulse structure ($I_{max}$), burst duration ($\tau_{dur}$), average pulse structure width ($<\tau_{wid}>$), and average interval between peaks ($<\tau_{intv}>$). The Table is segregated into the two burst groups with and without EE, and ordered by increasing $N_{pk}$. Clearly the non-EE group is dominated by bursts with a single pulse structure peak (22/41), for which there is no $<\tau_{intv}>$ entry; another quarter have two structures, but four non-EE bursts have 5–10 pulse structures. Whereas, half of the EE bursts have two structures; three EE bursts have 4–7 structures. Only one EE burst has a single structure. Figure 2 illustrates $I_{max}$ vs. $N_{pk}$ for the *Swift*/BAT short bursts. Filled diamonds are the 12 bursts with EE; open diamonds for 41 bursts without EE. There does not appear to be any evident trend connecting $I_{max}$ and $N_{pk}$ that might account for the preponderance of one-peak bursts in the non-EE group. However, the four non-EE bursts with 5–10 peaks are among the brightest of the whole sample; we return to these bursts when considering their $<\tau_{wid}>$ and $<\tau_{intv}>$.

The distributions of $N_{pk}$ for the two groups are shown in Figure 3. As the values contributing to the distributions are integers, it is not appropriate to use a Kolmogorov-Smirnov (K-S) test to estimate the probability that the two distributions are similar. Instead, we bootstrapped the larger (non-EE) distribution in Figure 3 to find how frequently the first two bins of the actual EE distribution were equaled, counting realizations where the remaining five $N_{pk}$ bootstrapped values were distributed anywhere within $3 \leq N_{pk} \leq 10$. Such configurations occurred $\sim 30$–40 times per run (100,000 realizations). Slightly different requirements (e.g., requiring $N_{pk=3} = 2$) are more constraining. We conclude that the probability for the actual EE $N_{pk}$ distribution to be drawn from a parent population like the non-EE one is $\lesssim 3.5 \times 10^{-5}$.

The temporal descriptors – $\tau_{dur}$, $\tau_{wid}$, and $\tau_{intv}$ – are essentially continuous variables, and so we use the K-S test to estimate probabilities that the EE and non-EE distributions are similar, as well as the stretch factors that render minimal K-S distances between each pair of distributions. The chance probabilities are listed in Table 2 along with the stretch factors for minimal K-S distance. We caution that the probabilities are not independent for at least two evident reasons. First, while pulse widths and intervals in a given burst may be independent (or not: there may be causation in either direction), duration is a combination of pulse widths and intervals. Second, we are analyzing the temporal descriptors in the observed frame. The non-EE (or EE) group may lie at preferentially higher redshifts; the resulting lower signal-to-noise levels would translate into less sensitivity for the higher redshift group in distinguishing both pulses and intervals.

Figure 4 illustrates the $\tau_{dur}$ distributions. The non-EE distribution is wider, nearly completely overlapping the EE distribution – as is the case with the $\tau_{wid}$ and $\tau_{intv}$ distributions, see below – and extending to significantly shorter durations. The K-S probability that the non-EE and EE distributions have similar shape is $\sim 6 \times 10^{-3}$. The stretch factor which renders their mutual K-S distance minimal is $\sim 3$. Thus the *Swift* EE durations tend to be longer, as presaged by the BATSE short burst durations (Figure 1).

Previous investigations have shown that pulse width and spectral lag are inversely correlated with pulse intensity (e.g., Norris 2002, Figure 2), that this anti-correlation exists within bursts, and unifies short and long bursts (Hakkila et al. 2008; Hakkila & Cumbee 2009; Hakkila & Preece 2011). In Figure 5 we plot pulse intensities versus widths. The previously identified trend is evident in the two short burst groups, with non-EE bursts extending to more intense, narrower pulses – while completely overlapping the low-intensity, wider-pulse region populated by most EE bursts. Open (filled) symbols designate non-EE (EE) bursts, with triangles (circles) for multi (single) pulse bursts. Figure 5b illustrates average pulse intensity versus average pulse width per burst. Non-EE (EE) bursts are indicated by open (filled) diamonds. The same trend is evident as in Figure 5a.

The pulse-structure width distributions – including all widths – are illustrated in Figure 6a. The same qualitative picture applies for $\tau_{wid}$ as for $\tau_{dur}$: the non-EE distribution is wider than the EE one, overlapping it and extending to significantly narrower pulses. The distributions in Figure 6a are dominated by bursts with several pulses. Plotted in Figure 6b are the distributions for average pulse width per burst ($<\tau_{wid}>$, the quantity recorded in Table 1). The K-S probability for similar-shaped $<\tau_{wid}>$ distributions is ~ 0.02, with stretch factor for minimal distance ~ 1.7.

Figure 7 illustrates pulse peak intensity versus pulse interval, $\tau_{intv}$. Open (filled) diamonds are for non-EE (EE) bursts. In (a) the intensity is that of preceding peak; in (b) that of succeeding peak. The picture is similar to that for pulse widths, with intervals in non-EE bursts extending to shorter timescales and associated with more intense peaks, and overlapping the region occupied by EE bursts. Preceding and succeeding peak intensities exhibit similar degrees of anti-correlation with pulse width. As the burst sources lie at different distances, some spread is expected in intensity, and in width due to time dilation. Since a burst must have $N_{pk} > 1$ to have at least one interval, Figure 7 does not include single-pulse bursts. Thus single-pulse non-EE bursts do not dominate the anti-correlation of pulse intensity versus general timescale, as can be seen by considering only the triangle symbols in Figure 5a. Possible imports of anti-correlations of pulse intensity with both width and interval are taken up in the discussion.

The distributions for all pulse intervals and $<\tau_{intv}>$ are shown in Figures 8a and 8b, respectively. The K-S probability for similar-shaped $<\tau_{intv}>$ distributions is ~ 0.002, with stretch factor for minimal distance ~ 2.4.

For non-EE bursts with several pulses ($N_{pk}$ = 5-10) – GRBs 051221A, 060313, 061201, and 090510 (see Table 1) – $<\tau_{wid}>$ and $<\tau_{intv}>$ are significantly shorter than for the EE bursts with similar $N_{pk}$. This is most likely at least partially a brightness bias effect as the $I_{max}$ for these four bursts are amongst the brightest for the *Swift* short burst sample, and so their temporal structure is more readily resolved.

In summary, there is a clear tendency for non-EE bursts to have one pulse, and EE bursts more than one. A previously identified trend unifying long and short bursts – the anti-correlation of pulse intensity and width (Hakkila & Preece 2011) – is manifest in the two short burst groups, with non-EE bursts extending to more intense, narrower pulses. Preceding and succeeding pulse intensity is also anti-correlated with pulse interval. The non-EE distributions for $\tau_{dur}$, $<\tau_{wid}>$, and $<\tau_{intv}>$ are wider, tend to smaller values, and have minimal-distance K-S stretch factors of ~ 3, 1.7, and 2.4, respectively, compared to the corresponding EE distributions.

## 4. XRT LIGHT CURVES OF SHORT BURSTS

As for long bursts, the initial fluxes and durations of X-ray afterglows of short bursts may be expected to be related to prompt emission fluence and density of environment, respectively. The initial steep X-ray decay phase often seen may be high-latitude emission arising from internal shocks; while the strength of later afterglow phases and thus the afterglow duration also depend on the environment within which external shocks are presumed to arise (O'Brien et al. 2006; Evans et al. 2009; Nakar 2007). Therefore we examined the *Swift*/XRT light curves of short bursts, available via the UK Swift Science Data Centre at the University of Leicester [4].

Construction of an XRT light curve (LC), while straightforward, requires application of scaling factors due to exposure and point-spread-function related considerations. The plotted {time, flux} points in a given LC may be derived from several snapshots of the GRB's afterglow, taken noncontiguously by the XRT in the course of observations scheduled for various other targets. Most salient for our purposes, each point in the LC contains a minimum of 15 detected photons (0.3–10 keV) and therefore its time extent increases as afterglow intensity drops. Details of XRT data and the LC construction process are elucidated in Evans et al. (2007).

Some of the short burst afterglows were not detected and so only estimates for flux upper limits (ULs) are available at the Gamma-ray Bursts Coordinates Network site in GCN circulars or reports [5]. Table 3 lists the flux ULs and exposure start and stop times for two EE and six non-EE bursts along with a one-point detection that was reported for GRB 090305A. No XRT observation was performed for GRBs 050202, 070923, and 090715A; no UL was reported for GRB 051105A.

Figure 9 illustrates the XRT fluxes (0.3–10 keV) at time of first detection ($F_{first}$) for the 49 bursts with at least one detection point or with an UL. Horizontal bars indicate the start and stop times for the first detection point, relative to the BAT trigger time. Filled (open) circles are EE (non-EE) burst detections; inverted triangles are ULs. Contrary to the more detailed calculation for the time value in the online XRT LCs, in Figure 9 we plot the symbol at the geometric mean of the start and stop times, since the distribution of XRT snapshots contributing to each LC point is somewhat arbitrary given the complex XRT observing schedule. For all but seven observations this center time is within 200 s of trigger, that is, during the initial steep decay phase lasting from XRT target acquisition (or even before) to ~ 500 s. It is apparent that the EE (non-EE) bursts tend to occupy the region of higher (lower) fluxes.

It is judicious to estimate what the observed distributions of $F_{first}$ would be if all were scaled to the same time after trigger. Because the initial decay phase is routinely steep, the $F_{first}$ for dimmer bursts span longer time intervals (recall, each flux point has a minimum of 15 photons), and their average flux values are thereby lowered. Assuming $F \propto T^{-3}$, roughly the average decay profile for the brighter bursts over the steep decay phase (during which most afterglow photons arrive, 80 s $\lesssim T \lesssim 500$ s), and using the start and stop times of the first detections, we scaled those fluxes in

---

[4] http://www.swift.ac.uk/xrt_curves/
[5] http://gcn.gsfc.nasa.gov/

Figure 9 (including ULs) with start times < 500 s to T = 100 s, the approximate median of the first detection times for the brighter bursts. Figure 10a illustrates the resulting $F_{T=100}$ distributions for EE bursts (solid histogram) and non-EE bursts (dashed histogram). The median $F_{T=100}$ for EE bursts is ~ $6 \times 10^{-10}$ erg cm$^{-2}$ s$^{-1}$, while the corresponding flux for non-EE bursts is ~ $3 \times 10^{-11}$ erg cm$^{-2}$ s$^{-1}$. Given that three ULs are included in the non-EE set (35 bursts), the included EE set (7 bursts) for Figure 10a has a median of $\gtrsim 20 \times$ brighter afterglow fluxes at T = 100 s post BAT trigger. While the EE set is small, accumulated by *Swift* over 2004–2010, Figure 10a suggests that the initial phases of EE short burst afterglows tend to be powered by a larger injection of energy than are non-EE afterglows. Interestingly, the X-ray fluxes of long bursts at T ~ 100 s in O'Brien et al. (2006, Figure 2) are comparable to the $F_{T=100}$ for EE bursts. But this similarity must be a coincidence arising from the disparate redshift distributions of long and short bursts being approximately compensated by their disparate total energy distributions (see Nysewander, Fruchter & Pe'er 2009 for related considerations).

The redshift distribution of short bursts as a whole is sparse, and several bursts whose optical positions lie outside the light profiles of candidate hosts contribute an additional measure of uncertainty. Also, short bursts with dim X-ray afterglows (or ULs) tend to have no detected optical afterglow and therefore association with the true host remains ambiguous, given the few arc second sizes of XRT error regions. Bursts with dim afterglows tend to have lower BAT fluences; these dimmer bursts may lie at higher redshift (for discussions of the complexity of the problem, see: Fong, Berger & Fox 2010; Berger 2010).

Thus we must consider that the $F_{T=100}$ distribution extending to lower fluences for non-EE bursts might be partially or wholly explained by the tendency of non-EE bursts to lie at higher redshift than EE bursts. The last column of Table 1 lists the secure and suggested redshifts for 19 of our sample of 53 bursts, which we have taken at face value from the UK Swift Science Data Centre archive. For the bursts without redshifts we assigned z = 0.5 and 1.0 for EE and non-EE bursts, respectively – approximately the median values for short bursts with redshifts. We then redshifted the flux values for the whole sample to a common frame of z = 0.75. We assumed an X-ray spectral slope of –1, for which unique slope $\nu L_\nu$ is independent of redshift and so the k-correction is moot (as noted by Nysewander et al. 2010; see Hogg 2000, eq. 22). The correction factor to z = 0.75 is then

$$F_{T=100,z=0.75} / F_{T=100} = (1 + 0.75) / (1 + z_{burst}) \times (d_{L,z=0.75} / d_{L,burst})^2 , \qquad (4)$$

where $d_L$ is luminosity distance and the ratio of (1 + z) factors accounts for the XRT finite bandpass. Figure 10b illustrates the resulting $F_{T=100, z=0.75}$ distributions, which tend to shift to lower (higher) values for EE (non-EE) bursts. The respective median fluxes ~ $4 \times 10^{-10}$ (~ 5 $\times$ $10^{-11}$) erg cm$^{-2}$ s$^{-1}$, are then only a factor of ~ 8 different. As Figure 10b is only a representative thought experiment – in which the unknown and insecure redshifts for short bursts could be much different than our assumptions – the tentative conclusion is that either non-EE bursts inject much less energy into their initial X-ray afterglow phase – believed attributable to high-latitude curvature radiation of the internal shock-driven prompt emission – than do EE bursts, and/or the two groups have significantly different redshift distributions.

Some illumination on the indeterminacy of redshift and prompt energy distributions may be had by examination of timescales. We also extracted from the XRT LCs the time of last detection ($T_{last}$) of the X-ray afterglow, or time of UL if no detection was made. Figure 11a illustrates the IPC durations ($\tau_{dur}$) vs. $T_{last}$ in the observed frame. The $\{\tau_{dur}$ (s), $T_{last}$ (s)$\}$ medians are $\{1.6, 6.4 \times 10^4\}_{EE}$ and $\{0.5, 2.2 \times 10^3\}_{non-EE}$. The $\tau_{dur}$ ratios are ~ 3.25, the $T_{last}$ ratios ~ 30. When the two timescales are shifted to the common frame with z = 0.75 frame using the same prescription as above for assigning redshifts, the result shown in Figure 11b is very similar: The medians become $\{1.3, 4.3 \times 10^4\}_{EE}$ and $\{0.4, 1.4 \times 10^3\}_{non-EE}$ with negligibly different $\tau_{dur}$ and $T_{last}$ ratios compared to those in the observed frame.

Thus different redshift distributions cannot account for the disparate $\tau_{dur}$ and $T_{last}$ ratios. Medians as different as $z_{EE}$ ~ 0.5 and $z_{non-EE}$ ~ 1.0 for the two thirds of short bursts with unknown redshifts are supportable (but not required; see considerations on likely redshift distributions of short bursts in Berger 2010), but more disparate redshift distributions for the two groups would only yield even larger $\tau_{dur}$ and $T_{last}$ ratios. The *Swift* sample of EE bursts is small, but the similar trend seen in the BATSE sample (Figure 1; also Konus sample – see discussion) for EE bursts to have durations on the long end of the short burst duration distribution argues that the $\tau_{dur}$ ratio is real. Longer EE burst durations (~ × 3) combined with wider pulses (~ × 2, Figure 6) still falls short of the $T_{last}$ ratio ~ 30, suggesting that total energy difference is insufficient to account for the longer X-ray afterglows of EE bursts. The remaining factor determining external-shock dominated, later afterglow fluxes in standard GRB theory is density ($n^{\frac{1}{2}}$) of environment into which the burst energy is injected. Density could be lower on average for non-EE bursts: Nysewander, Fruchter & Pe'er (2009) inferred that environment densities for long and short bursts may be comparable, or even higher for short bursts. However, we note that their analysis of optical and X-ray afterglows necessarily excluded X-ray ULs, as well as afterglows not detected after 11 hrs (~ 40,000 s), a canonical reference time. In fact 28 (38 total) of the non-EE bursts have detections with $T_{last}$ < 40,000 s, or only an UL; whereas 7 of 11 EE bursts were detected with $T_{last}$ > 40,000 s (Figure 11a). Hence without explicit countervailing evidence, we conclude non-EE bursts could explode into lower density environments than do EE bursts, as Figure 11 may suggest.

# 5. DISCUSSION

In NGS10 we presented evidence of absence of the extended emission component for a portion of the *Swift*/BAT sample of short bursts: The BAT has sufficient sensitivity to detect EE in ~ half the sample, but so far had detected EE in only 12 of 53 short bursts. This conclusion of NGS10 left open whether an ancillary mechanism with a threshold operates below which EE is not manifest, or if EE and non-EE bursts arise from fundamentally different mechanisms. If the latter possibility were true, then evidence for different progenitors might be expected at some level during the initial, intense pulse complexes and/or during the X-ray afterglows. (If more than one progenitor, then other differences in "prompt emissions", e.g. in neutrino or gravitational wave channels, might also be expected to be found in future experiments.) We have described here three different kinds of evidence – besides the EE itself – that suggest more than one kind of progenitor may give rise to short bursts.

First, the characteristic timescales (durations, pulse-structure widths and intervals) for the IPCs of non-EE bursts tend to be shorter by factors of ~ 2–3 than those of EE bursts, and non-EE bursts tend to have fewer pulse structures (Figures 2–8). Even though the BAT sample of EE bursts is still relatively small, this finding was presaged by the BATSE duration distribution for short bursts (Figure 1). The Konus 1995–2000 sample (Mazets et al. 2002) also evidences that EE bursts occupy the long end of the short-burst duration distribution (visual estimate for median duration ~ 0.5 s: GRBs 951014A, 980605, 980706, 981107, 990313, 990327, 990516, 990712A, 000218, and 000701B).

In the duration distributions for all three experiments (*Compton*/BATSE, Konus, *Swift*/BAT) the non-EE durations extend to the upper end, overlapping the longest EE durations. As the NGS10 result only constrained that EE could have been detected in ~ half of the BAT sample, this allows that, if present, the EE component could have been too dim to be detected in the other half. This possibility of a larger EE fraction is consistent with the non-EE and EE distributions overlapping for pulse widths and intervals, as well as for X-ray fluxes and times of last detection.

Thus the distributions of all characteristic timescales, as well as initial X-ray fluxes and X-ray afterglow durations, that we have measured could in principle be more disjoint for the two short burst groups, but more sensitive experiments would be required to make such discriminating separations of EE and non-EE bursts. More disjoint timescales for EE and non-EE bursts would illuminate theoretical models for different merger types (NS-NS, NS-BH, WD-WD, WD-NS, AIC WD) discussed for short bursts (Eichler et al. 1989; Meszaros & Rees 1992; Narayan et al. 1992, Paczynski 1991; Mochkovitch et al. 1993; Metzer et al. 2008; 2010).

We also find that the anti-correlation of pulse intensity and width (Hakkila et al. 2008; Hakkila & Cumbee 2009; Hakkila & Preece 2011) – a unifying trend spanning long and short bursts – continues in the two short burst groups, with non-EE bursts extending to more intense, narrower pulses (Figure 5). Preceding and succeeding pulse intensities are also anti-correlated with pulse interval (Figure 6). Whatever process makes for the tendency of shorter timescales in non-EE bursts – e.g., more closely timed accretion events onto the incipient black hole, from perhaps smaller radii – concomitantly governs pulse emission strength, width and interval.

Second, for EE bursts the median of fluxes (corrected to T=100 s post trigger) for the initial, steep X-ray afterglow phase (80 s $\lesssim$ T $\lesssim$ 500 s) – thought to arise from the high-latitude curvature radiation of internal shock-driven prompt emission – is higher by factor of ~ 20 than for non-EE bursts (Figure 10). When we correct fluxes to a common redshift, assuming $z_{non-EE} = 1.0$ and $z_{EE} = 0.5$ for those without determined redshifts, the difference in median fluxes is only a factor of ~ 8. We note that the energy injection from the EE component can sometimes dominate the total prompt emission, and may account in some cases for the brighter initial X-ray afterglows of short bursts (e.g., Norris & Bonnell 2006; Perley et al. 2009); viewing angle and beaming factor may influence the ratio of IPC:EE flux and as well as prompt emission to initial afterglow strength (Metzger 2008).

Using a smaller sample of short bursts Sakamoto & Gehrels (2009) originally suggested the possibility that non-EE bursts have dimmer afterglows – as per above – and that their afterglows may be shorter-lived, our last point (Figure 11). In the X-ray afterglow phases succeeding the often-present plateau phase (T $\gtrsim$ few $\times$ $10^4$ s), the observed flux levels may depend on prompt-energy injection, redshift, and density of environment (Nakar 2007). Assuming a higher redshift distribution for non-EE bursts only slightly increases the ratio of times of last detection (Figures 11a & 11b). In addition to the often higher energy input from the EE component, the fact that X-ray afterglows of EE bursts tend to last ~ 30 $\times$ longer than non-EE afterglows might be partially accounted for by a higher density environment. Modeling indicates that dynamically-formed compact binaries in globular clusters dominate merger rates relevant to short GRBs (Lee, Ramirez-Ruiz & van de Ven 2010). These mergers may tend to occur while still within the globular cluster where densities are much higher than in the intergalactic medium (Parsons, Ramirez-Ruiz & Lee 2009).

In conclusion, the tendency for bursts without EE toward shorter temporal structures may reflect a coalescence process with significantly different dynamical timescale compared to the EE bursts with longer timescales. The longer-lasting X-ray afterglows of EE bursts suggests that a significant fraction explode into more tenuous environments than non-EE bursts, or that the sometimes-dominant EE component efficiently powers the afterglow. Combined, our results favor different progenitors for EE and non-EE short bursts.

We would like to thank Jon Hakkila, the referee, for illuminating conversations and for pointing out the anti-correlation between pulse intensity and pulse width inherent in the data set. We are grateful for suggestions from Alexei Pozanenko and Enrico Ramirez-Ruiz, and for conversations with Brian Metzger.

This work made use of data supplied by the UK *Swift* Science Data Centre at the University of Leicester and the US *Swift* Data Archive at NASA/Goddard Space Flight Center.

REFERENCES

Barthelmy, S.D., et al.  2005, Nature, 438, 994

Berger, E.  2009, ApJ, 690, 231

Berger, E.  2010, ApJ, 722, 1946

Eichler, D., Livio, M., Piran, T., & Schramm, D. N.  1989, Nature, 340, 126

Evans, P.T., et al.  2009, MNRAS, 397, 1177

Fenimore, E., et al.  2003, in AIP 662, Gamma-Ray Burst and Afterglow Astronomy 2001, ed. G. Ricker & R. Vanderspek, p. 491

Fong, W., Berger, E., & Fox, D.B.  2010, ApJ, 708, 9

Frederiks, D. D., et al. 2004, in ASP 312, Gamma-Ray Bursts in the Afterglow Era, ed. M. Feroci et al. (ASP: San Francisco), p. 197

Gehrels, N., et al.  2006, Nature, 444, 1044

Hakkila, J., Giblin, T.W., Norris, J.P., Fragile, P.C., & Bonnell, J.T.  2008, ApJ, 677, L81

Hakkila, J., & Cumbee, R.S. 2009, in AIP 1133, Gamma-Ray Bursts: Sixth Huntsville Symposium, eds. C. Meegan, N. Gehrels, & C. Kouveliotou (New York: AIP), p. 379

Hakkila, J., & Preece, R.D.  2011, ApJ (submitted; arXiv:1103.5434)

Hogg, D.W.  2000, arXiv:astro-ph\9905116v4

Jackson, B., et al.  2003, arXiv:math/0309285

Kann, D.A., et al.  2008, ApJ submitted (arXiv/ 0804.1959)

Kouveliotou, C., et al.  1993, ApJ, 322, L21

Lee, W.H., Ramirez-Ruiz, E., van de Ven, G.  2010, ApJ, 720, L953

Mazets, E.P., et al.  1981, Ap&SS, 80, 3

Mazets, E. P., et al. 2002, Konus Catalog of Short GRBs (St. Petersburg: Ioffe LEA), http://www.ioffe.ru/LEA /shortGRBs/Catalog/

Meegan, C. A., et al. 1996, in AIP Conf. Proc. 384, Gamma-Ray Bursts, ed. C. Kouveliotou, M. F. Briggs, & G. J. Fishman (New York: AIP), p. 291

Meszaros, P., & Rees, M. J. 1992, ApJ, 397, 570

Metzger, B.D., Quataert, E., & Thompson, T.A.  2008, MNRAS, 385, 1455

Metzger, B.D., Arcones, A., Quataert, E., & Martínez-Pinedo, G.  2008, MNRAS, 402, 2771

Mochkovitch, R., Hernanz, M., Isern, J., & Martin, X.  1993, Nature, 361, 236

Minaev, P.Yu., Pozanenko, A.S., & Loznikov, V.M.  2010, Astr. Letters, 36, 707

Nakar, E.  2007, Physics Reports, 442, 166

Narayan, R., Paczýnski, B., & Piran, T. 1992, ApJL, 395, L83

Norris, J.P., Cline, T.L., Desai, U.D., & Teegarden, B.J.  1984, Nature, 308, 434

Norris, J.P., Scargle, J. D., & Bonnell, J. T. 2001, in Gamma-Ray Bursts in the Afterglow Era, ed. E. Costa, F. Frontera, & J. Hjorth (Berlin: Springer), 40

Norris, J.P. 2002, ApJ, 579, 386

Norris, J.P., & Bonnell, J.T. 2006, ApJ, 643, 266

Norris, J. P., & Gehrels, N. 2008, AIP Conf. Proc. 1000, Gamma-ray Bursts 2007 (Melville, NY: AIP), 280 (NB06)

Norris, J.P., Gehrels, N., & Scargle, J.D. 2010, ApJ, 717, 411 (NGS10)

Nysewander, M., Fruchter, A.S., & Pe'er, A. 2009, ApJ, 701, 824

O'Brien, P.T., et al. 2006, ApJ, 647, 1213

Paczynski, B. 1991, Acta Astronomica, 41, 257

Perley, D.A., et al. 2009, ApJ, 696, 1871

Parsons, R.K., Ramirez-Ruiz, E., Lee, W.H. 2009 (submitted to ApJ; arXiv:0904.1768)

Sakamoto, T., & Gehrels, N. 2009, AIP 1133, Gamma-Ray Bursts: Sixth Huntsville Symposium, p. 112

Scargle, J. D. 1998, ApJ, 504, 405

Scargle, J.D., Norris, J.P., Jackson, B., & Chiang, J. 2011, in preparation

Troja, E., King, A.R., O'Brien, P.T., Lyons, N., & Cusumono, G. 2008, MNRAS, 385, L10

Villasenor, J.S., et al. 2005, Nature, 437, 855

TABLE 1.  Temporal Descriptors of BAT Short Burst IPCs

| GRB Date | $N_{pk}$ | $I_{max}$ (Counts s$^{-1}$) | $\tau_{dur}$ (s) | $<\tau_{wid}>$ (s) | $<\tau_{intv}>$ (s) | z |
|---|---|---|---|---|---|---|
| Extended Emission: | | | | | | |
| 080503 | 1 | 34 | 140 | 140 | ... | ... |
| 050911 | 2 | 17 | 1545 | 370 | 1042 | 0.165 |
| 061210 | 2 | 535 | 55 | 19 | 32 | 0.41 |
| 061006 | 2 | 216 | 420 | 141 | 280 | 0.4377 |
| 090916 | 2 | 17 | 3900 | 788 | 3013 | ... |
| 050724 | 2 | 59 | 1235 | 230 | 1022 | 0.258 |
| 090715 | 2 | 54 | 335 | 160 | 158 | ... |
| 051227 | 3 | 29 | 1940 | 109 | 764 | ... |
| 090531 | 3 | 43 | 2030 | 309 | 841 | ... |
| 071227 | 4 | 29 | 1625 | 162 | 390 | 0.394 |
| 060614 | 7 | 62 | 5360 | 295 | 813 | 0.125 |
| 070714 | 7 | 136 | 2745 | 175 | 385 | 0.923 |
| No Extended Emission: | | | | | | |
| 070923 | 1 | 149 | 45 | 45 | ... | ... |
| 050509B | 1 | 36 | 35 | 35 | ... | 0.226 |
| 050813 | 1 | 5 | 560 | 560 | ... | 1.25 [a] |
| 050925 | 1 | 63 | 120 | 120 | ... | ... |
| 051105A | 1 | 54 | 55 | 55 | ... | ... |
| 050202 | 1 | 32 | 220 | 220 | ... | ... |
| 100816A | 1 | 59 | 1965 | 1965 | ... | 0.804 |
| 090621B | 1 | 102 | 95 | 95 | ... | ... |
| 090815C | 1 | 5 | 695 | 695 | ... | ... |
| 060801 | 1 | 64 | 110 | 110 | ... | 1.131 |
| 100117A | 1 | 42 | 170 | 170 | ... | ... |
| 061217 | 1 | 21 | 355 | 355 | ... | 0.827 |
| 070209 | 1 | 42 | 70 | 70 | ... | ... |
| 091109B | 1 | 192 | 100 | 100 | ... | ... |
| 070714A | 1 | 21 | 1240 | 1240 | ... | ... |
| 070724A | 1 | 10 | 450 | 450 | ... | 0.457 |
| 070729 | 1 | 11 | 975 | 975 | ... | ... |
| 090515 | 1 | 143 | 15 | 15 | ... | ... |
| 070809 | 1 | 11 | 805 | 805 | ... | ... |
| 100206A | 1 | 140 | 135 | 135 | ... | 0.41 |
| 090426 | 1 | 9 | 1465 | 1465 | ... | 2.6 |
| 081226A | 1 | 17 | 235 | 235 | ... | ... |
| 080702A | 2 | 22 | 695 | 127 | 530 | ... |
| 060502B | 2 | 102 | 475 | 72 | 385 | 0.287 |
| 081024A | 2 | 58 | 1790 | 92 | 1700 | ... |
| 100702A | 2 | 62 | 525 | 120 | 342 | ... |
| 090305A | 2 | 19 | 340 | 104 | 216 | ... |
| 100625A | 2 | 123 | 270 | 113 | 122 | ... |
| 070731 | 2 | 7 | 2540 | 850 | 1838 | ... |
| 051210 | 2 | 10 | 1495 | 632 | 519 | 1.4 |
| 080919 | 2 | 22 | 760 | 352 | 274 | ... |
| 070810B | 2 | 56 | 125 | 19 | 86 | ... |
| 071112B | 2 | 30 | 740 | 84 | 667 | ... |
| 070429B | 3 | 27 | 560 | 81 | 216 | 0.902 |
| 090607 | 3 | 18 | 2880 | 146 | 1211 | ... |
| 080905A | 3 | 92 | 950 | 117 | 364 | ... |
| 100213A | 3 | 44 | 2065 | 208 | 958 | ... |
| 090510 | 5 | 210 | 865 | 41 | 143 | 0.903 |
| 051221A | 6 | 719 | 1030 | 37 | 66 | 0.5465 |
| 061201 | 6 | 174 | 615 | 39 | 91 | 0.111 |
| 060313 | 10 | 379 | 765 | 33 | 61 | ... |

[a] average of two reported redshifts:  0.7 and 1.8

TABLE 2.  Comparison of Temporal Distributions:  Short Bursts with and without EE

| | $N_{pk}$ | $\tau_{dur}$ | $\tau_{wid}$ [a] | $\langle\tau_{wid}\rangle$ [b] | $\tau_{intv}$ [a] | $\langle\tau_{intv}\rangle$ [b] |
|---|---|---|---|---|---|---|
| Probability | $3.5\times10^{-5}$ | $6.2\times10^{-3}$ | $2.0\times10^{-5}$ | $1.8\times10^{-2}$ | $4.5\times10^{-5}$ | $2.2\times10^{-3}$ |
| Stretch Factor | ... | 2.98 | 2.22 | 1.72 | 3.12 | 2.43 |

[a] all pulses (or intervals)

[b] average pulse (or interval) per burst

TABLE 3.  Short Burst XRT Upper Limits

| GRB Date | Flux (erg cm$^{-2}$ s$^{-1}$) | $T_{min}$ (s) | $T_{max}$ (s) | Reference |
|---|---|---|---|---|
| **Extended Emission:** | | | | |
| 050911 | $< 3\times10^{-14}$ | $1.7\times10^{4}$ | $4.6\times10^{4}$ | GCN Circ. 3976, Page et al. |
| 090916 | $< 3\times10^{-14}$ | $3.4\times10^{4}$ | $1.3\times10^{5}$ | GCN Rep. 248.1, Troja et al. |
| **No Extended Emission:** | | | | |
| 051105A | $< 3\times10^{-14}$ | $10^{2}$ | $2.8\times10^{4}$ | GCN Circ. 4034, Holland et al. |
| | | | | GCN Circ. 4043, Beardmore, et al. |
| 070209 | $< 3\times10^{-14}$ | $8.9\times10^{1}$ | $6.5\times10^{4}$ | GCN Circ. 6095, Stratta et al. |
| | | | | GCN Circ. 6119, Stratta et al. |
| 070731 | $< 10^{-14}$ | $1.4\times10^{5}$ | $3.0\times10^{5}$ | GCN Rep. 79.1, Moretti et al. |
| 070810B | $< 5\times10^{-14}$ | $7.0\times10^{1}$ | $1.4\times10^{4}$ | GCN Rep. 81.1, Marshall et al. |
| 071112B | $< 2\times10^{-13}$ | $3.7\times10^{3}$ | $5.8\times10^{3}$ | GCN Rep. 103.1, Perri et al. |
| 090305A [a] | $5.5\times10^{-12}$ | $10^{2}$ | $1.9\times10^{2}$ | GCN Circ. 8937, Beardmore, et al. |
| 090815C | $< 7\times10^{-14}$ | $2.2\times10^{4}$ | $6.5\times10^{4}$ | GCN Circ. 9824, Cummings et al. |

[a] One-point detection for GRB 090305A.

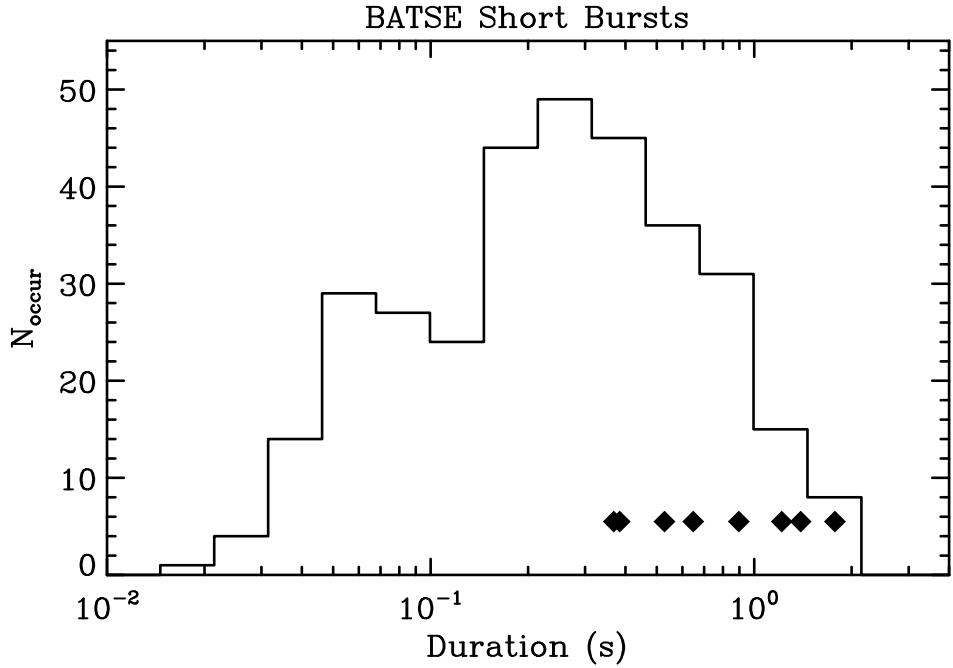

Fig. 1 – Duration distribution for 327 BATSE short bursts where the TTE data cover the whole IPC. Duration measure is comparable to $T_{90}$ measure (see text). Diamonds are the eight BATSE short bursts with extended emission discussed in Norris & Bonnell (2006).

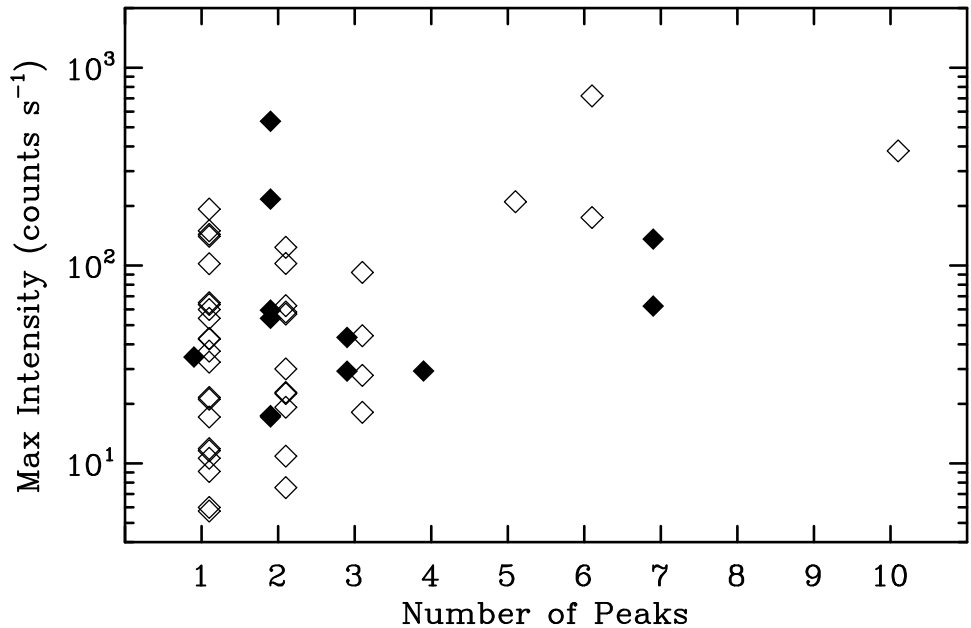

Fig. 2 – Intensity ($I_{max}$) of the brightest pulse structure above background level vs. number of peaks ($N_{pk}$) for the *Swift*/BAT short bursts. Filled diamonds are the 12 bursts with EE; open diamonds for 41 bursts without EE. Symbols are displaced slightly, left and right, for clarity.

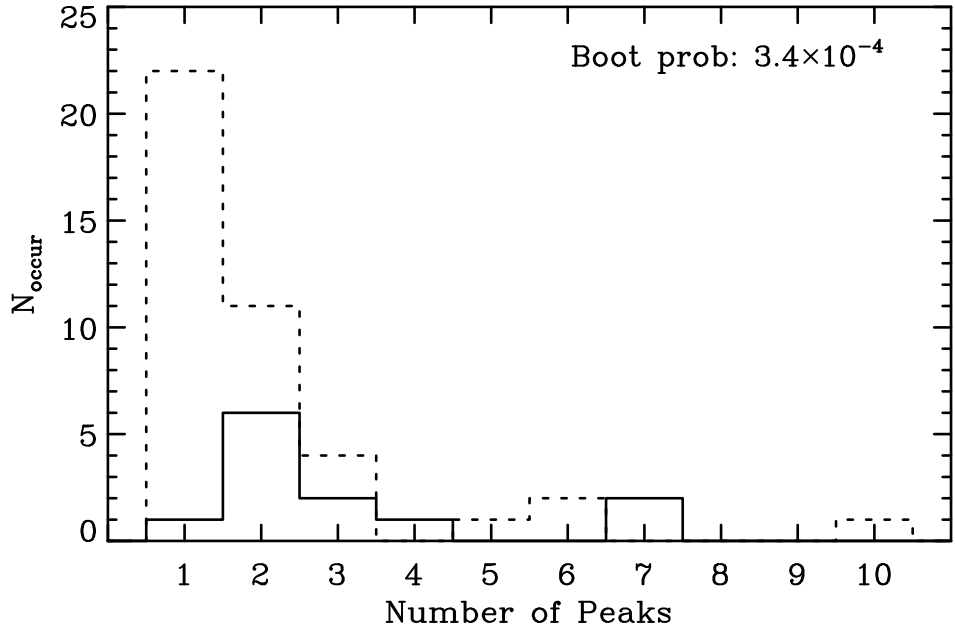

Fig. 3 – Distributions of $N_{pk}$ for bursts with EE (solid histogram) and bursts without EE (dashed histogram). Annotations for K-S probability and stretch factor described in text for this and following figures.

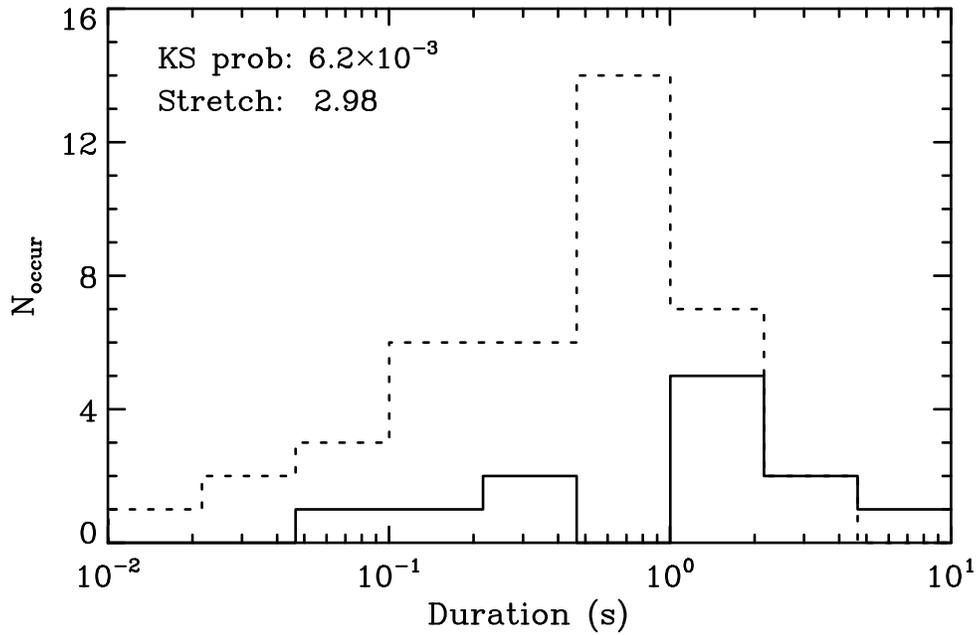

Fig. 4 – Distributions of $\tau_{dur}$, IPC duration, for bursts with EE (solid histogram) and bursts without EE (dashed).

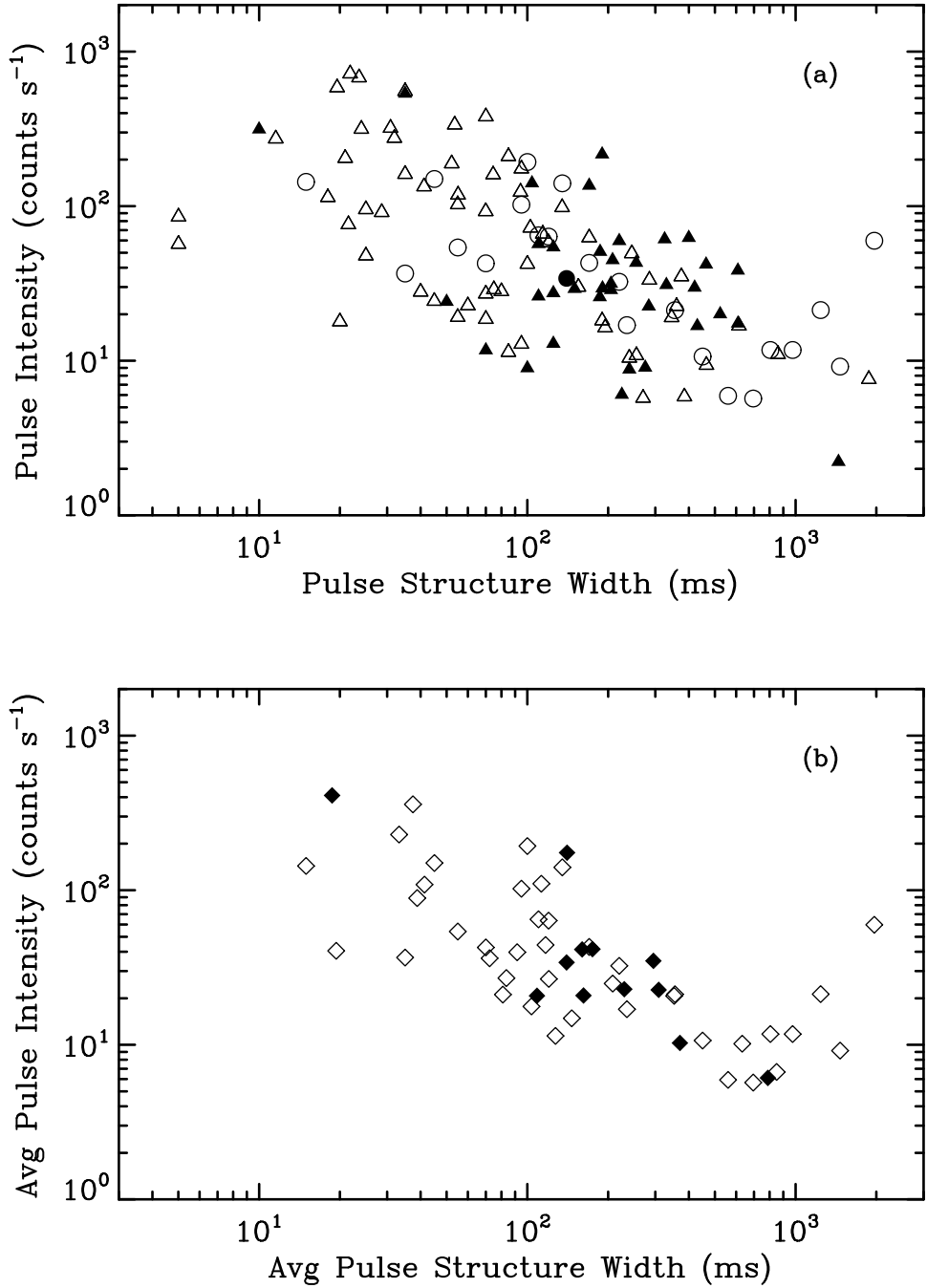

Fig. 5 – (a) Pulse intensity vs. pulse structure width, $\tau_{\mathrm{wid}}$. Open (filled) symbols for non-EE (EE) bursts, with triangles (circles) for multi (single) pulse bursts. (b) Average pulse intensity and width per burst for non-EE (EE) indicated by open (filled) diamonds.

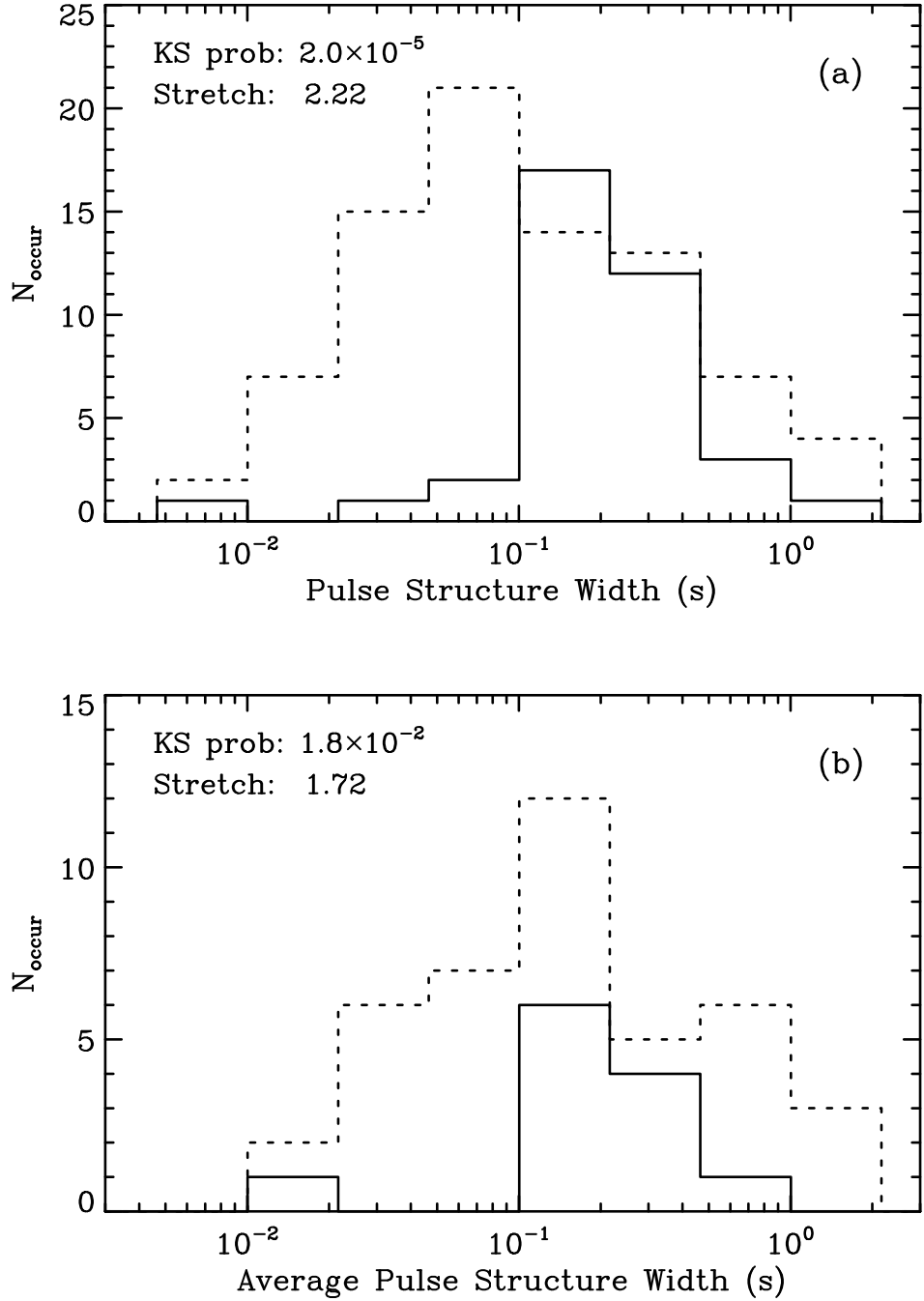

Fig. 6 – (a) Distributions of $\tau_{wid}$, pulse structure width, for bursts with EE (solid histogram, 37 structures) and bursts without EE (dashed, 83 structures). (b) Distributions of the average $\tau_{wid}$ per burst for the two groups.

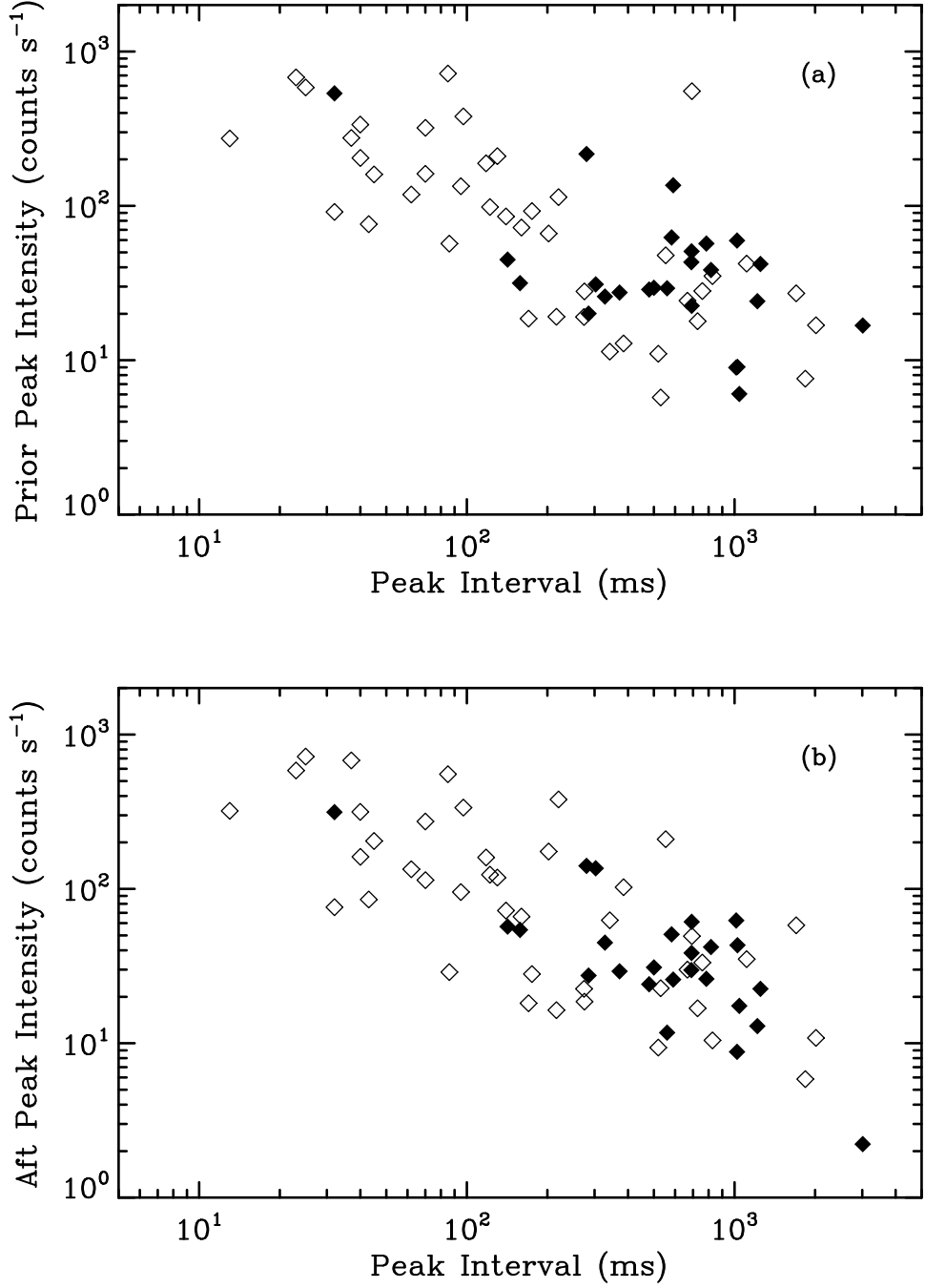

Fig. 7 – Pulse peak intensity vs. pulse interval, $\tau_{intv}$; open (filled) diamonds for non-EE (EE) bursts. In (a) the intensity is that of preceding peak; in (b) that of succeeding peak.

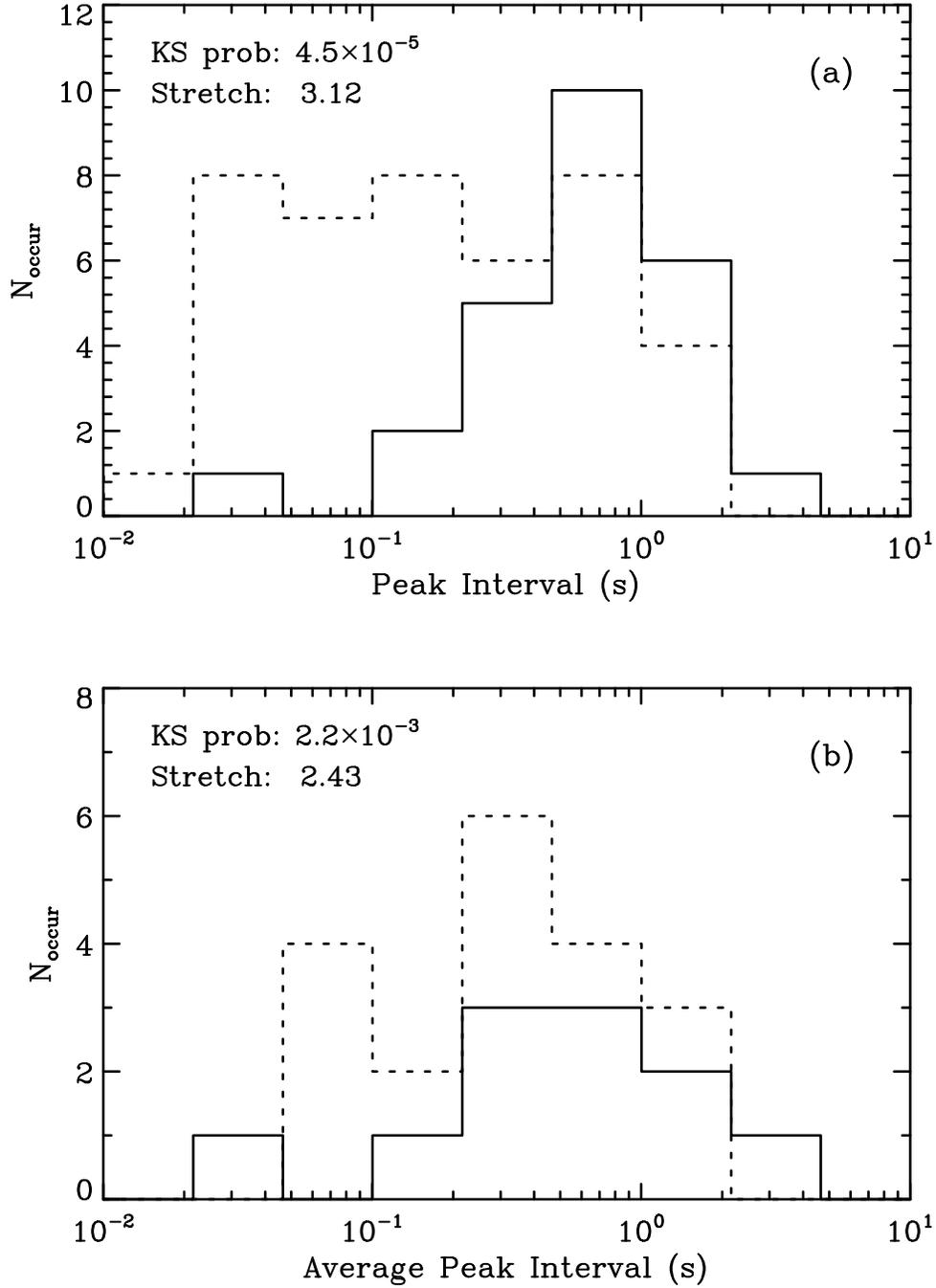

Fig. 8 – (a) Distributions of $\tau_{intv}$, interval between two peaks, for bursts with EE (solid histogram, 25 intervals) and bursts without EE (dashed, 42 intervals). (b) Distributions of the average $\tau_{intv}$ per burst for the two groups.

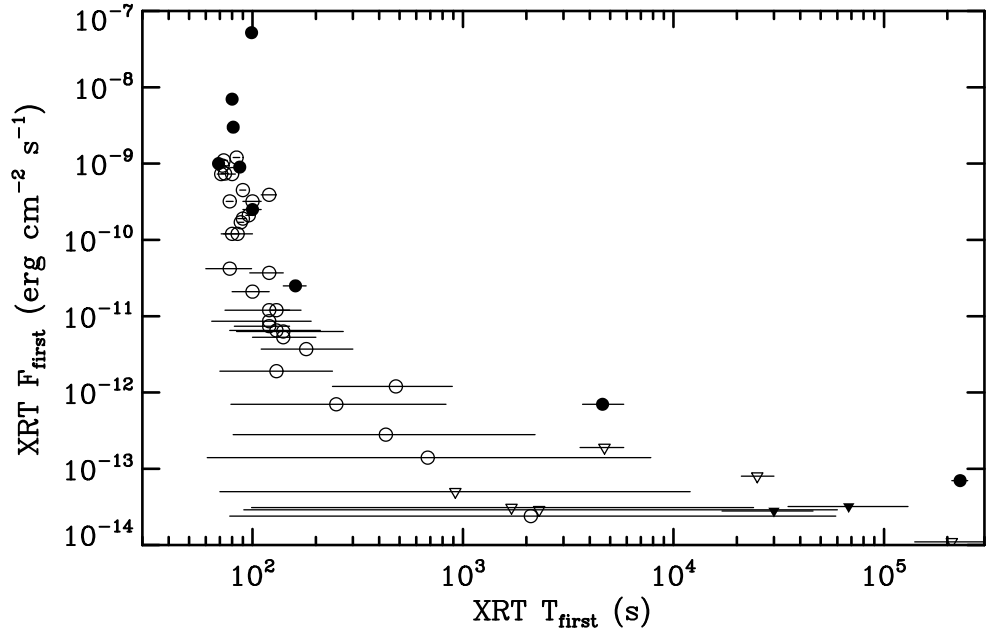

Fig. 9 – XRT flux at time of first detection for 49 short bursts with at least one detection point or an upper limit. Horizontal bars indicate start and stop times of XRT snapshots contributing to data point, relative to BAT trigger time. Filled (open) circles are EE (non-EE) bursts; inverted triangles are ULs.

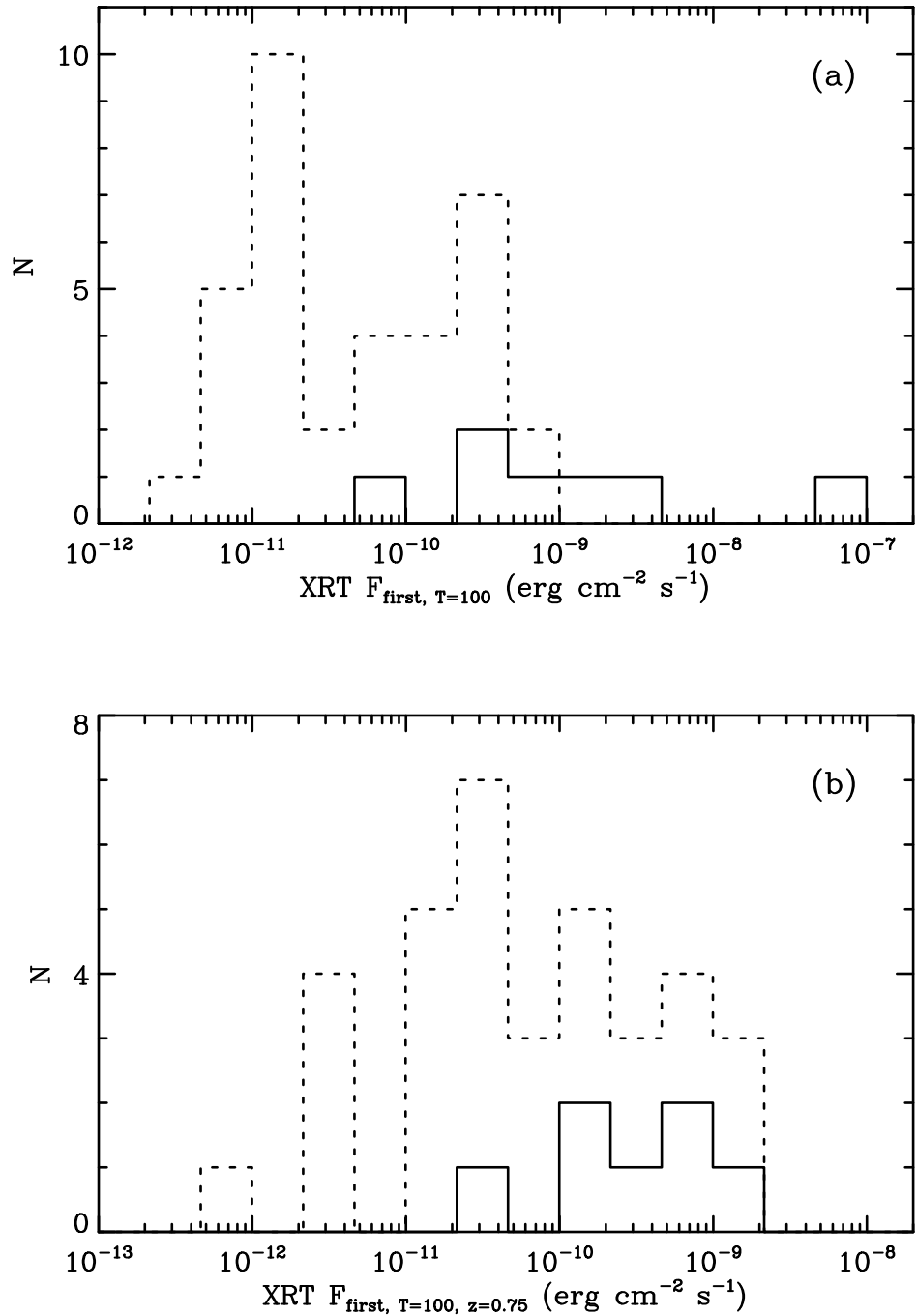

Fig. 10 – Distributions of XRT flux at time of first detection (or upper limit) for bursts with integration start time < 500 s, with flux scaled ($F \propto T^{-3}$) to T = 100 s relative to BAT trigger. (a) Flux in observer frame. (b) Flux shifted to common frame (z = 0.75) using published redshifts, or assuming $z_{EE}$ ($z_{non-EE}$) = 0.5 (1.0) for bursts without redshifts. Note that flux scale for (b) is shifted one decade lower than in (a).

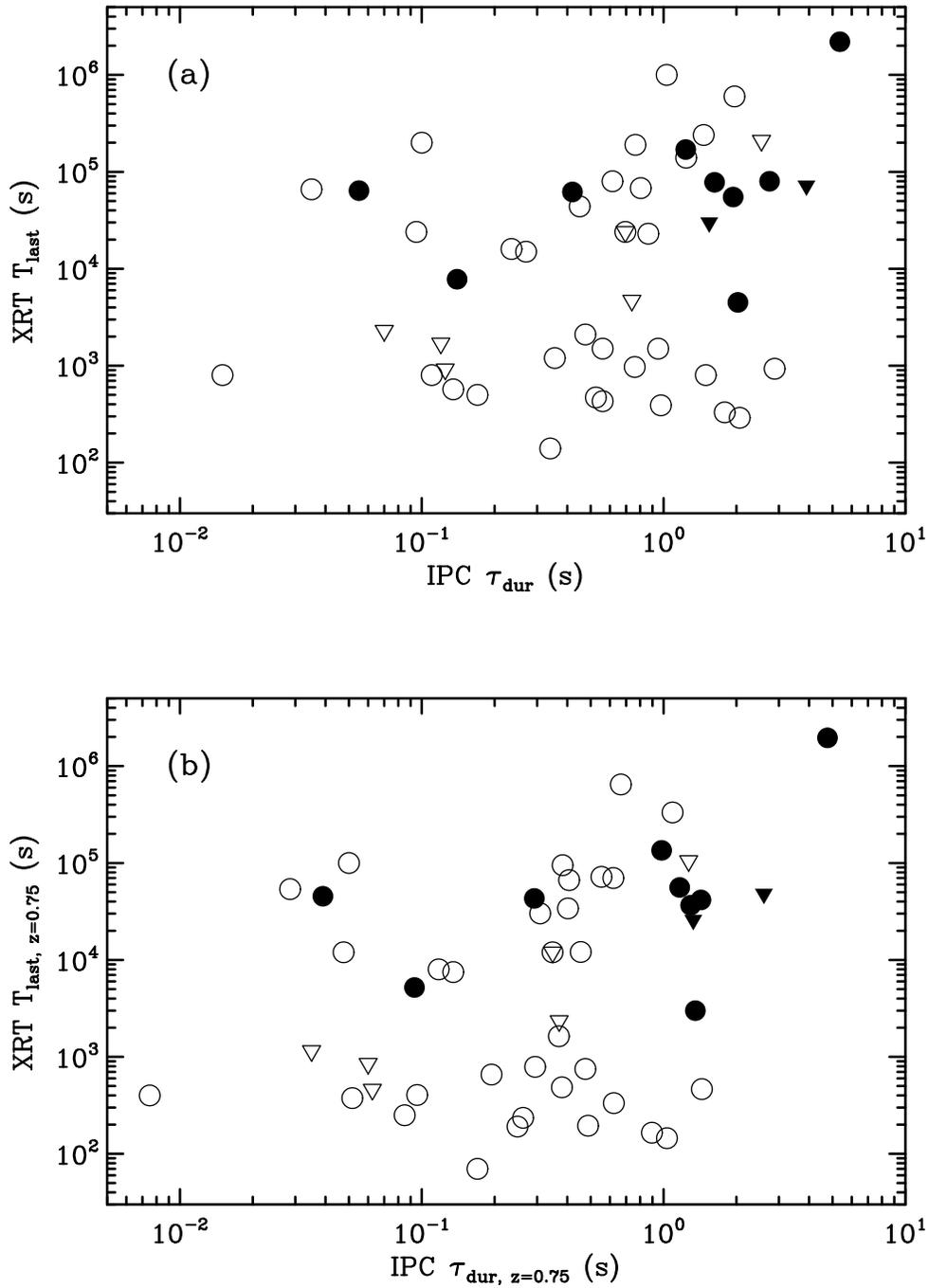

Fig. 11 – XRT time of last detection (or upper limit) relative to BAT trigger versus IPC duration. (a) In observer frame. (b) Both time intervals shifted by $(1+z)^{-1}$ to common frame ($z = 0.75$) using published redshifts, or assuming $z_{EE}$ ($z_{non-EE}$) = 0.5 (1.0) for bursts without redshifts.